\documentclass[aps,prb,twocolumn,10pt,longbibliography]{revtex4-1}
\usepackage{amsmath}
\usepackage{amsfonts}
\usepackage{amssymb}
\usepackage{mathtools}
\usepackage{hyperref}
\usepackage{wasysym}
\usepackage{bm}
\usepackage{cases}
\usepackage[version=4]{mhchem}
\usepackage{graphicx}
\usepackage[caption=false]{subfig}
\allowdisplaybreaks%

\begin{document}
\title{Intrinsic transverse field in frustrated quantum Ising magnets:\\ 
its physical origin and quantum effects}
\author{Gang Chen}
\affiliation{Department of Physics and Center of Theoretical and Computational Physics, 
the University of Hong Kong, Hong Kong, China}
 
\date{\today}    

\begin{abstract}
Transverse field Ising model is a common model in quantum magnetism and 
is often illustrated as an example for quantum phase transition. Its physical 
origin in quantum magnets, however, is actually not quite well-understood. 
The quantum mechanical properties of this model on frustrated systems are not 
well-understood either. We here clarify the physical origin, both extrinsic 
one and intrinsic one, for the transverse field of the quantum Ising model, 
and then explain the quantum effects in the Kagome system. 
We discuss the quantum plaquette order and the quantum 
phase transition out of this ordered state in the rare-earth Kagome magnets. 
Our specific results can find their relevance in the rare-earth tripod 
Kagome magnets.  
\end{abstract}

\maketitle

\section{Introduction}

The classical Ising model is a textbook model in the field of magnetism 
and statistical physics. The exact solution by Lars Onsager for the two-dimensional
Ising model is a milestone of modern statistical physics and proved the 
very existence of continuous phase transitions with only short-range 
interactions~\cite{PhysRev.65.117}. The far-reaching impact of Onsager's 
solution goes much beyond the original motivation~\cite{PhysRev.76.1232,PhysRev.76.1244}. 
Its quantum extension, the transverse field Ising model, contains the ingredient
of the quantum phase transition and emergent low-energy quantum field theories
at the criticality~\cite{SubirBook}. For the unfrustrated Ising interaction on systems 
like a square lattice, the tranverse field Ising model can be well-understood from 
the high-dimensional classical Ising model and its thermal transition. On the 
frustrated systems, however, new ingredients may arise from the interplay 
between the quantum fluctuation and the geometrical frustration
of the underlying lattices.

Besides the interesting physical properties of the transverse field Ising models, the 
physical origin of the transverse field Ising models is actually not well understood.
This is related to the physical realization of this simple and important model. 
The Ising model requires a strong spin anisotropy in the spin space, and this 
almost immediately implies that, the magnetic system must have a strong 
spin-orbit coupling. Indeed, the localized moments of the proposed Ising magnets,
such as quasi-$1d$ magnets CoNb$_2$O$_6$, BaCo$_2$V$_2$O$_8$, SrCo$_2$V$_2$O$_8$ and 
various 2d/3d rare-earth magnets, do arise from the strong spin-orbit entanglement, 
and the local moments have a strong orbital character~\cite{CoNbO,SrCoVO,FaureBCVO,
PhysRevB.96.024439,PhysRevLett.120.207205,2018nature25466,2019arXiv190305492W,2019arXiv190411413C}. 
In the case of the Co$^{2+}$
local moment, the ion has a $3d^7$ electron configuration and has one hole 
in the lower $t_{2g}$ shell, and the spin-orbit coupling is active here. 
As a result, the ion has a total spin ${S=3/2}$ and an effective orbital 
angular momentum ${L=1}$, and the resulting total moment is given by the 
spin-orbit-entangled Kramers doublet. Because of the involvement of the  
orbital degrees of freedom, the exchange interaction between the Kramers 
doublet has to be anisotropic. This is indeed the underlying driving force
for the Kitaev interactions in the Co-based honeycomb magnets Na$_2$Co$_2$TeO$_6$ 
and Na$_3$Co$_2$SbO$_6$~\cite{PhysRevB.97.014407,PhysRevB.97.014408}, 
and the anisotropic interaction in the pyrochlore 
cobaltate NaCaCo$_2$F$_7$~\cite{Krizan2015,PhysRevMaterials.1.074412,PhysRevB.95.144414}. 
For the case of quasi-1d magnets CoNb$_2$O$_6$, 
BaCo$_2$V$_2$O$_8$ and SrCo$_2$V$_2$O$_8$, because of the local Co$^{2+}$ 
environment and the special lattice geometry, the system realizes the Ising 
interactions between the local moments. The transverse field is then introduced
externally by applying a magnetic field normal to the Ising spin direction. 
This is feasible because the Co$^{2+}$ local moment is a Kramers doublet and 
all the three components of the moments are magnetic. This is the {\sl external} 
origin of the transverse field.

\begin{figure}[b]
\centering
\includegraphics[width=7cm]{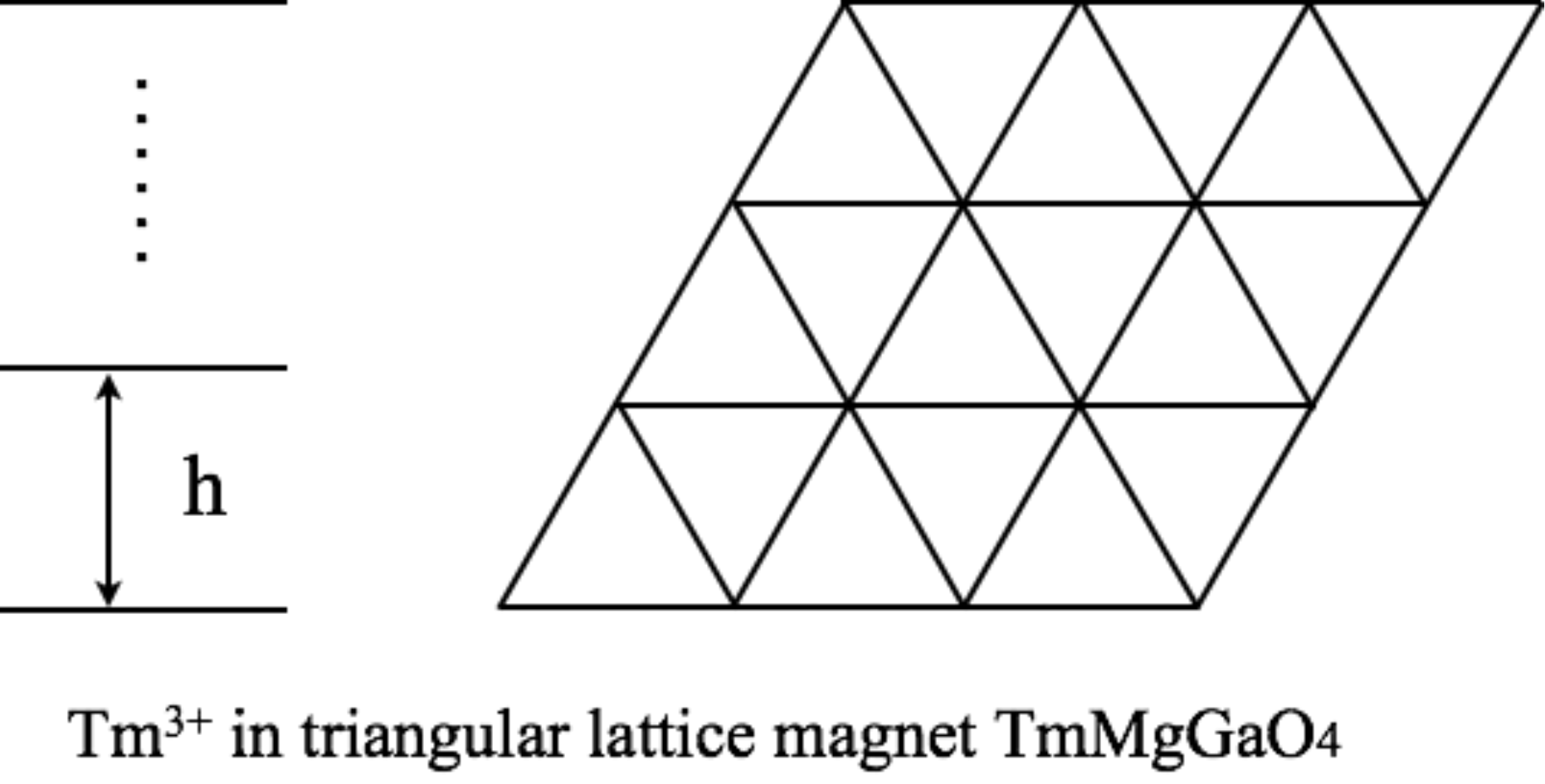}
\caption{The Tm$^{3+}$ magnetic ions in TmMgGaO$_4$ form
a triangular lattice~\cite{CavaTMGO,YueshengTMGO,YaoshenTMGO}. 
The lowest two crystal field singlets can be modelled as an 
effective spin-1/2 degree of freedom, and the weak crystal field
splitting is modelled as a transverse field. This aspect of 
microscopics and physical model have been clarified in 
Ref.~\onlinecite{YaoshenTMGO}.} 
\label{fig1}
\end{figure}

Is there an {\sl intrinsic} origin of the transverse field? Our successful modelling~\cite{YaoshenTMGO}
of the intertwined multipolar physics in the triangular lattice magnet TmMgGaO$_4$~\cite{CavaTMGO,YueshengTMGO}
suggests a positive answer. We start from our early understanding about 
the Tm$^{3+}$ ion in TmMgGaO$_4$ and then give an answer for the general cases. 
The $4f$ electrons of the Tm$^{3+}$ ion has a total spin ${S=1}$ and orbital angular 
momentum ${L=5}$, then the spin-orbit coupling leads to a total moment ${J=6}$. 
As we show in Fig.~\ref{fig1}, the two lowest crystal field states of the Tm$^{3+}$ ion 
are two singlets and they comprise two independent singlet irreducible representations
of the D$_{3d}$ point group rather than one single two-dimensional irreducible 
representation. Because the total moment, $J$, is integer in nature, 
there is no Kramers theorem, and there is always a crystal field 
energy gap separating these two singlets~\cite{YaoshenTMGO}. 
This differs fundamentally from the usual non-Kramers doublet
whose degeneracy is protected by the two-dimensional irreducible representation 
of the D$_{3d}$ point group. The two singlet wavefunctions primarily involve 
the ${J^z=\pm 6}$ components, and the group representation forbids ${J^z=\pm 1}$ 
here~\cite{YaoshenTMGO}. If one defines a pseudospin-1/2 degree of freedom that operates 
on these two singlets, only the $z$ component of the pseudospin carries the magnetic dipole 
moment, and the remaining two components are high order multipole moments that are related to 
high order products of $J$ moments. Because of this very unique property, the external magnetic 
field only acts on the $z$ component of the pseudospin, regardless of the direction 
of the external magnetic field. Thus the transverse field cannot be generated externally. 
Remarkably, our mother nature builds the transverse field intrinsically in this 
material~\cite{YaoshenTMGO} and brings quantum phenomena within itself. 
The observation is from the weak crystal field splitting that was modelled 
by us as the transverse field on the transverse multipolar component of the 
pseudospin~\cite{YaoshenTMGO}.

What can we learn from the successful and special example of TmMgGaO$_4$?
The two weakly separated crystal field singlets of the Tm$^{3+}$ ions are the consequences 
of the singlet representations of the point group and the non-Kramers nature of 
the ion. More generally, the point group symmetry of other rare-earth magnets
is not as high as the D$_{3d}$ point group, and thus it is more common to have
the singlet crystal field states for non-Kramers ions with integer spins. Therefore, 
in the rare-earth magnets with a low point group symmetry, if the two lowest 
singlets are well separated from the higher excited crystal field states, then we can 
single them out and build up a pseudospin-1/2 degree of freedom. This pseudospin-1/2 
degrees of freedom will be responsible for the low-temperature magnetic properties. 
Like our case for the Tm$^{3+}$ ion, the longitudinal component of the pseudospin 
is the dipole-like, and the transverse component is the multipole-like. 
If the terms of the wavefunctions for the two singlets are 
far apart in the $J^z$ basis, then the exchange interaction will be mostly 
Ising-like because it is a bit difficult for the system to flip ``$J^z$'' multiple 
times during the superexchange process. More importantly, the remaining splitting 
between the two singlets could be modelled as the transverse field on the multipole 
component~\cite{YaoshenTMGO}.

The above discussion clarifies an intrinsic origin for the transverse 
field of the transverse field Ising model that can potentially be relevant
for many rare-earth magnets. To further explore the quantum 
properties on the frustrated system, we turn to a specific frustrated 
lattice that is Kagome lattice. Apparently, the tripod rare-earth Kagome
magnets have already existed~\cite{PhysRevLett.116.157201,PhysRevB.95.104439,
2016NatComm713842P,PhysRevB.98.174404}. Moreover, the rare-earth Kagome magnets 
can be obtained from the rare-earth pyrochlore magnets from the 
dimensional reduction by applying an external magnetic field along 
the $[111]$ crystallographic direction that polarizes one 
sublattice~\cite{2017arXiv171202418L,2007NatPhys566F,Matsuhira2002}.  
For a Kagome lattice, the 2-fold degenerate non-Kramers doublet
is not allowed by the lattice symmetry, and there should always
be a splitting between two lowest crystal field states. This 
naturally provides the transverse field. The major parts of this paper
are to probe the existence and explore the experimental consequences 
of this intrinsic transverse field. Due to the strong geometrical 
frustration, the pure transverse field Ising model with an 
antiferromagnetic Ising interaction on the Kagome lattice 
is known to have no transition throughout all parameter regime~\cite{PhysRevB.71.024401}, 
and the system is smoothly connected to the polarized state in 
the strong transverse field limit. To create more structures 
from the geometrical frustration and the transverse
field, we apply an external magnetic field to the system 
and establish a phase diagram with a phase transition. We show that, at the weak 
transverse field regime, the system develops a quantum plaquette
order. There can be a direct quantum phase transition 
from this quantum plaquette order to the disordered state. 
We explain the dynamical properties in different regime 
and discuss some key experimental consequences.

The remaining parts of the paper are organized as follows. 
In Sec.~\ref{sec2}, we introduce our physical model
and justify our introduction of the external magnetic field. 
In Sec.~\ref{sec3}, we carry out the perturbative treatment in the Ising limit
and establish the quantum plaquette order.
In Sec.~\ref{sec4}, we regard the quantum plaquette ordered state
as the parent state and demonstrate the relation between this state
and the disordered state. We show that, the quantum plaquette order can be
regarded as a confining phase of the compact U(1) gauge theory and  
the disordered state can be regarded as the Higgs phase. 
In Sec.~\ref{sec5}, we explain the thermodynamic properties and 
establish the excitations inside the disordered phase,
and this excitation spectrum would be a strong evidence for the intrinsic 
transverse field in the system and the multipolar structure 
of the local moments. 
Finally, in Sec.~\ref{sec6}, we conclude with a discussion of 
the experimental relevance and consequences.

\section{Kagome lattice transverse field Ising model} 
\label{sec2}

We start with a brief introduction of the rare-earth tripod Kagome magnet
A$_2$RE$_3$Sb$_3$O$_{14}$ where A = Mg, Zn and RE refers to the rare-earth
atom (Pr, Nd, Gd, Tb, Dy, Ho, Er, Yb, Tm)~\cite{PhysRevLett.116.157201,PhysRevB.95.104439,
2016NatComm713842P,PhysRevB.98.174404}. Various interesting phases 
and results have already been suggested for this new family of materials. 
The Nd$^{3+}$, Dy$^{3+}$, Er$^{3+}$, and Yb$^{3+}$ ions have odd number 
of electrons per ion and thus support a Kramers doublet locally. For 
these Kramers doublet local moments, if the Ising spin is realized,
the transverse field has to be generated externally. The Gd$^{3+}$ ion is special 
and has a half-filled $4f$ shell with a total spin ${S=7/2}$ and a quenched
orbital angular momentum, the atomic spin-orbit coupling is inactive for this ion. 
The remaining ones all have integer total moments. 
Unlike the rare-earth pyrochlore magnets where non-Kramers doublets exist in 
many compounds, there is no such non-Kramers doublet in the tripod Kagome magnets. 
The symmetry of the Kagome lattice is too low to support a 2-fold non-Kramers
degeneracy. There is always a finite crystal field splitting between the 
would-be non-Kramers doublet. Similar to the context of the pyrochlore magnets,
we can still introduce an effective spin-1/2 degree of freedom here except 
that we need to introduce a transverse field to take care of the crystal 
field splitting between the two singlets of the non-Kramers doublets.
Like the case for the Tm$^{3+}$ ion in TmMgGaO$_4$, the transverse field 
is of intrinsic origin. The resulting model is given as
\begin{eqnarray}
H = \sum_{ij} [J_{ij} S^z_i S^z_j + \cdots] - \sum_i  h S^x_i ,
\end{eqnarray}
where ``$\cdots$'' refers to the XY-like spin flipping term ($S^+_i S^-_j$)
and the pair-flipping term ($S^+_i S^+_j$) that can be written down from 
the symmetry analysis, and these extra terms would have a strong
bond dependence and bring the quantum fluctuations to the Ising part.
This kind of anisotropic spin model has been widely studied in the 
context of the rare-earth pyrochlores and the triangular lattice spin 
liquid materials. This anistropic interaction on the Kagome lattice 
has not yet been discussed in the literature and we will address it 
in an another paper. Here, since the intrinsic transverse field 
already brings quantum properties into the system, we will focus 
on this transverse field Ising model. From the materials' point of view, 
we have learned from the experience of the rare-earth pyrochlore magnets
that some materials such as Ho$_2$Ti$_2$O$_7$ do behave quite Ising-like~\cite{Fennell}. 
It is natural to expect that, in the tripod Kagome system, this Ising 
feature could persist.  Indeed, it was proposed that the Dy-based
and Ho-based tripod Kagome magnets do behave Ising-like where the 
intrinsic transverse field was indicated for the Ho-based one~\cite{2018arXiv180604081D}.  

The ferromagnetic quantum Ising model does not lead to any unknown properties
even on this frustrated lattice. Thus, we consider an antiferromagnetic 
Ising interaction and restrict ourselves to the nearest neighbors. It was 
actually studied numerically long time ago that the transverse field Ising 
model on the Kagome lattice has neither phase transition nor symmetry breaking. 
The system remains disordered throughout the parameter space~\cite{PhysRevB.71.024401}. 
This was referred as ``disorder-by-disorder''~\cite{PhysRevB.71.024401}, 
in contrast to the ``order-by-disorder''~\cite{PhysRevLett.84.4457,
PhysRevB.63.224401}
for the transverse field Ising model on the 
triangular lattice in the weak field regime~\cite{PhysRevLett.84.4457,
PhysRevB.63.224401,PhysRevB.93.235103,PhysRevB.97.085114}. 
The essential reason for the 
disordered state even in the weak field regime arises from the fact that
the up-up-down and down-down-up triangular plaquettes are degenerate and 
both of them appear in the low energy manifold. To create more structures
to the phase diagram, we here apply an external magnetic field to the system. 
Only the dipolar component ($S^z$) of the pseudospin will couple to the external
magnetic field regardless to the orientation of the external magnetic field. 
This field could have the effect of removing half of the active
spin configurations in each triangular plaquette. So our model now becomes
\begin{eqnarray}
H = \sum_{\langle ij \rangle} J S^z_i S^z_j -  \sum_i  h S^x_i 
                                            -  \sum_i  B S^z_i ,
\end{eqnarray}
where only the nearest-neighbor interaction is considered. 
In the absence of the intrinsic transverse field $h$, an infinitesimal
external magnetic field $B$ would already select all up-up-down spin configuration
and create a magnetization plateau. This magnetization plateau persists 
up to the field value of $J/2$ without extra interactions. This plateau 
regime in $B\in (0,J/2)$ is identical to the ``Kagome spin ice'' that was obtained from the 
classical pyrochlore spin ice by applying a magnetic field along the 
[111] direction to polarize one sublattice~\cite{2017arXiv171202418L,
2007NatPhys566F,PhysRevB.68.064411}. 
However, this is classical physics. In the following section, we analyze 
the quantum effect of the intrinsic transverse field within this 
degenerate spin manifold. Some aspects of this model such as the ordered 
structure have been established in the early numerical 
study~\cite{PhysRevLett.84.4457,PhysRevB.63.224401}. 
Our purpose is to propose this model for the rare-earth Kagome magnets with 
non-Kramers doublets. We further propose the fractionalized nature of the 
phase transition based on the lattice gauge theory, and explore the experimental 
consequences of different phases and the transition due to the multipolar 
nature of the local moments. The experimental signature of the intrinsic 
transverse field is emphasized.

\section{Perturbation theory and quantum plaquette orders} 
\label{sec3}

\begin{figure}[t]
\centering
\includegraphics[width=8cm]{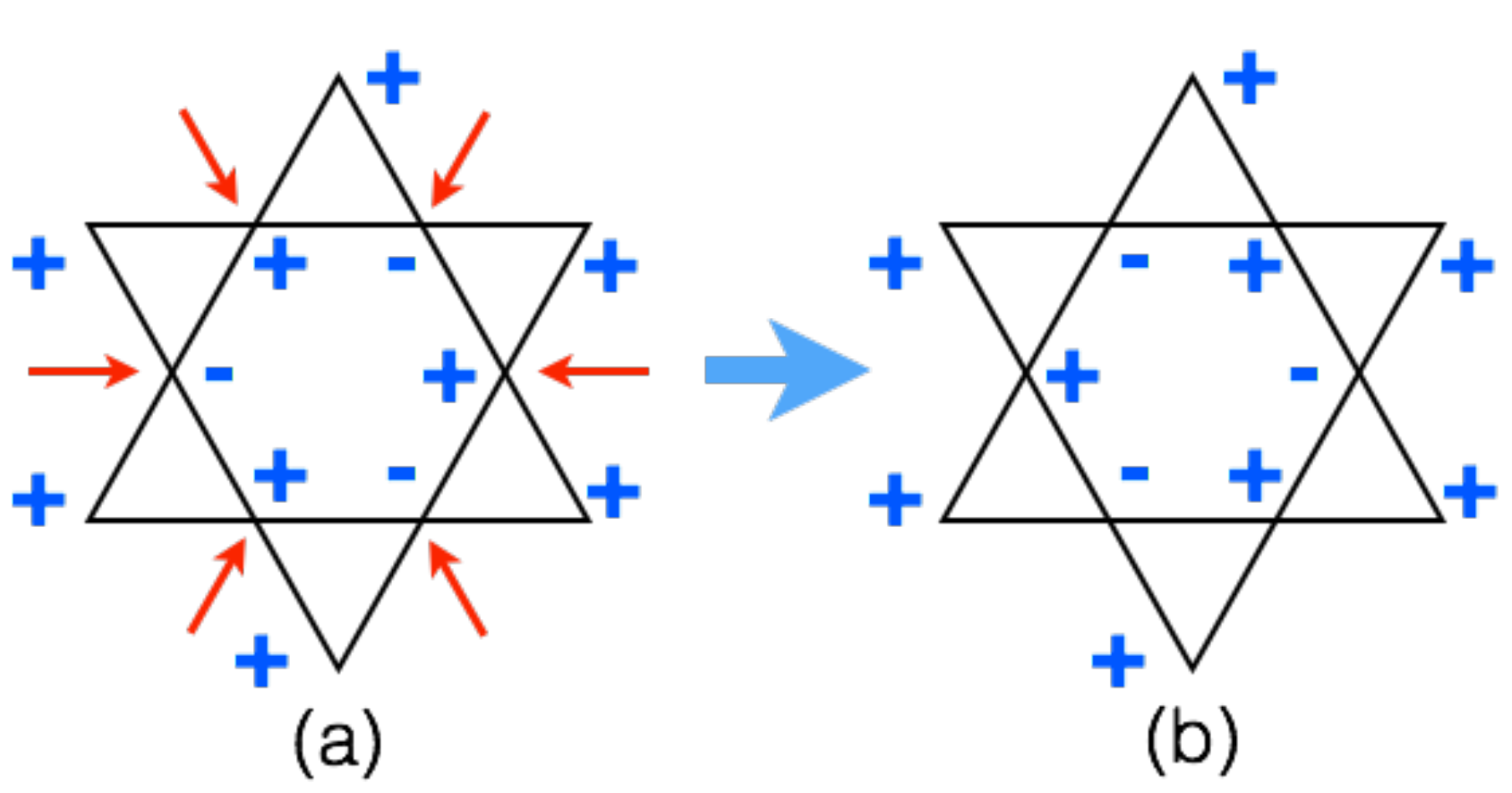}
\caption{The sixth order degenerate perturbation process from the intrinsic transverse field 
within the degenerate ``up-up-down'' spin configurations. The red arrows indicate
the application of the transverse field on this site. ``+'' refers to spin-up,
and ``$-$'' refers to spin-down. See the main text for the detailed discussion.} 
\label{fig2}
\end{figure}

\begin{figure}[b]
\centering
\includegraphics[width=6cm]{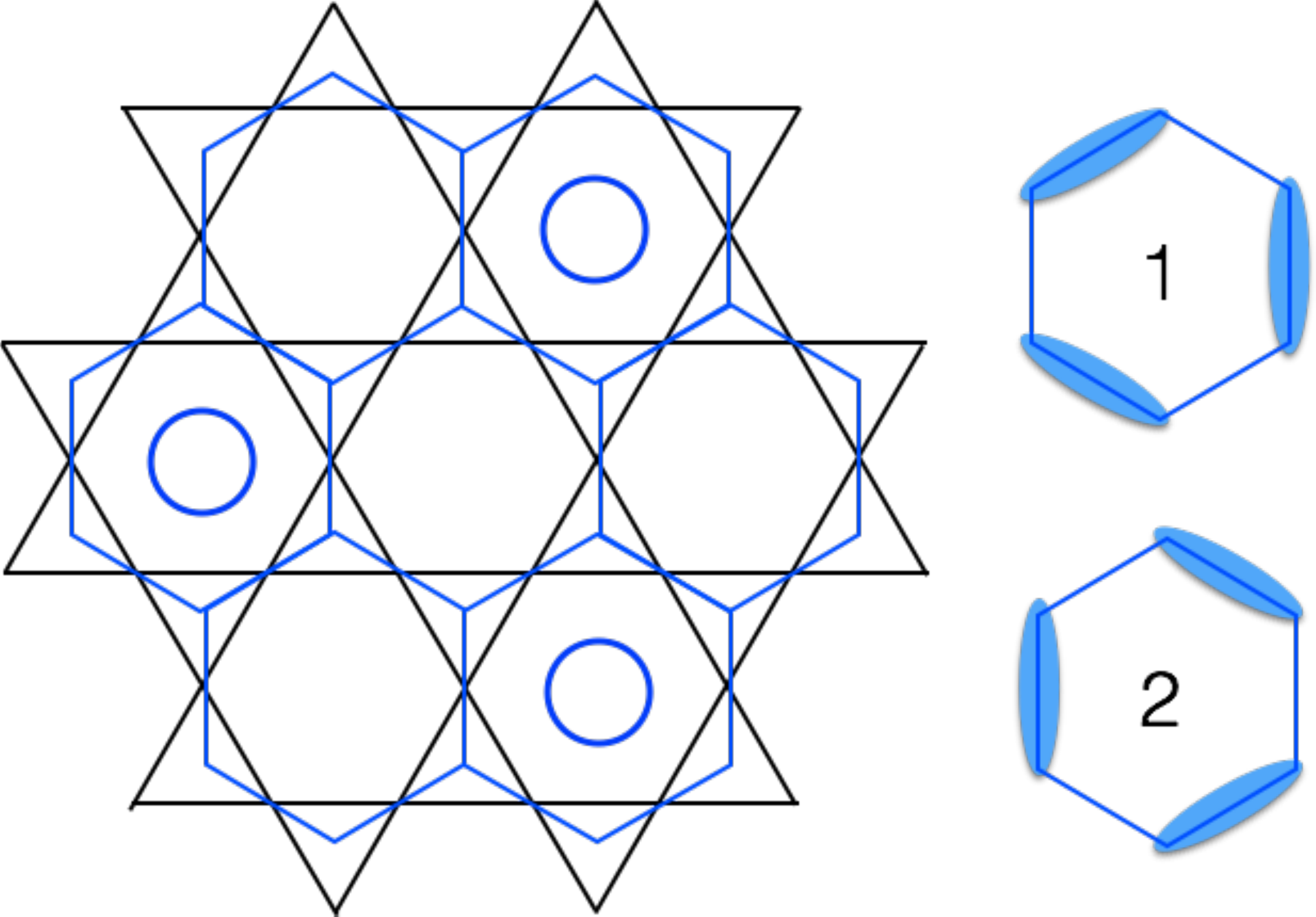}
\caption{The Kagome lattice and its dual honeycomb lattice. Dimer configurations 1 and 2 
refer to two distinct dimer coverings on the elementary hexagon. In the quantum plaquette 
ordered phase, the hexagons with blue circles refer to the resonating hexagon where the
three dimers form a quantum linear superposition of the dimer configurations 1 and 2. } 
\label{fig3}
\end{figure}

\begin{figure}[t]
\centering
\includegraphics[width=8cm]{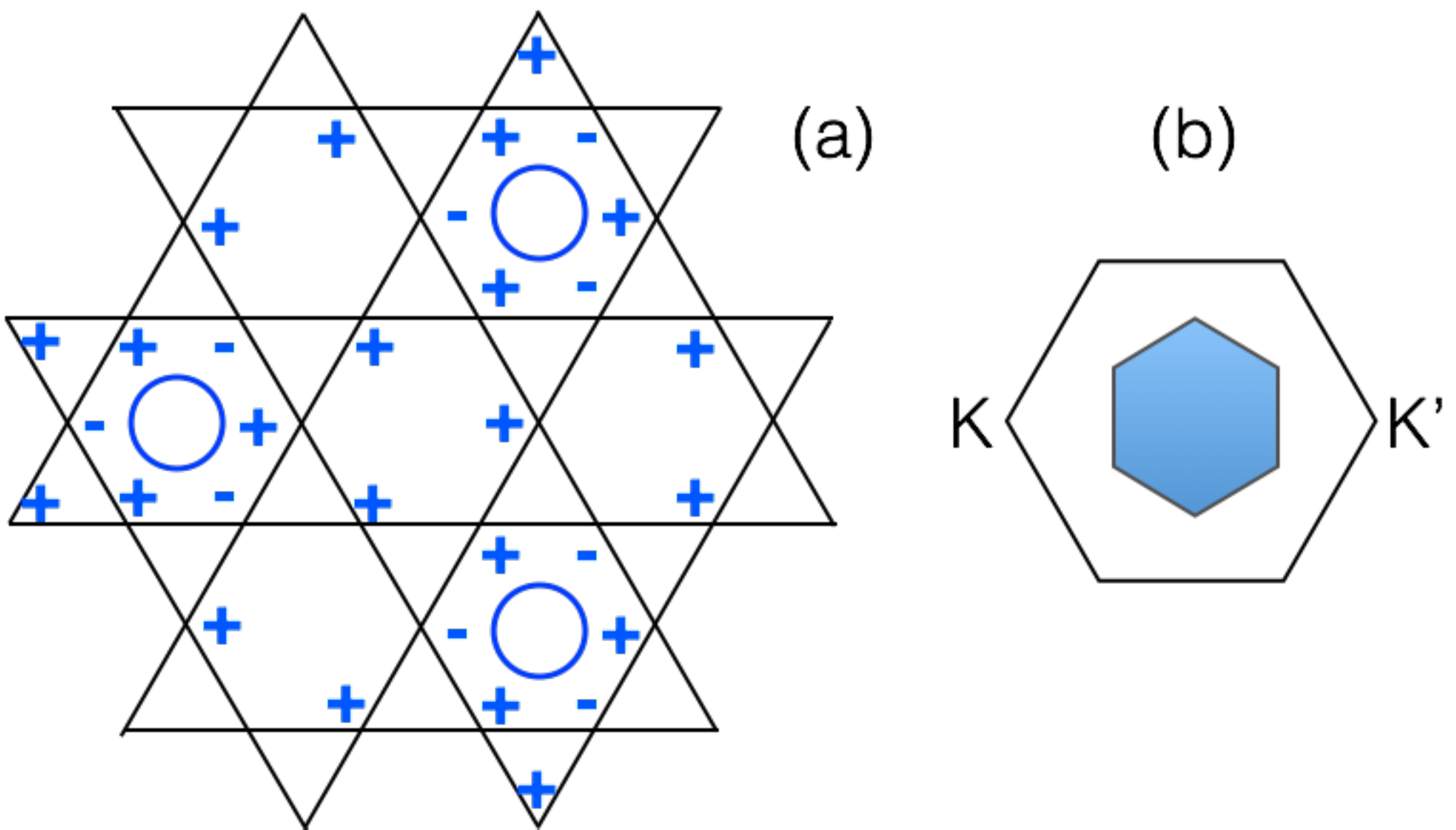}
\caption{The quantum plaquette order in terms of the spin configuration. On the resonating 
hexagon, the six spins form a quantum linear superposition of the alternating spin orientations.
The ordering wavevectors of the quantum plaquette order occur at K and K$'$ points. 
The blue hexagon inside the big hexagon in (b) is the reduced Brillouin zone when
the system develops the quantum plaquette order. 
} 
\label{fig4}
\end{figure}

We continue to work within the degenerate ``up-up-down'' spin configurations due
to our introduction of the external magnetic field on the dipolar component of the 
pseudospin. Once a weak transverse field is introduced, the extensive degeneracy 
will be lifted, and a degenerate perturbation theory is needed. The leading effect 
comes from the sixth order, that is depicted in Fig.~\ref{fig2}, is summarized below, 
\begin{eqnarray}
H_{6} = - J_6 \sum_{\hexagon} S^x_1 S^x_2 S^x_3 S^x_4 S^x_5 S^x_6,
\end{eqnarray}
where ${J_6 \sim{\mathcal{O}(h^6/J^5)}>0}$ and the ``$-$'' sign takes care of 
even number of perturbation series. 
The lower order perturbations either vanish or give a constant shift to the 
classical energies. Our sixth order effective Hamiltonian, 
$H_6$, operates on the degenerate manifold of ``up-up-down'' spin configuration.  
This process can be mapped to a quantum dimer model on the dual honeycomb 
lattice that is formed by connecting the centers of the triangular plaquettes 
if the down spin is mapped to the dimer convering on the bond connecting the 
centers of the neighboring triangular plaquettes (see Fig.~\ref{fig3}). 
This is an exact mapping. The quantum dimer model is given as 
\begin{eqnarray}
H_6 =- J_6 \sum_{\varhexagon} [ |\varhexagon_1 \rangle \langle \varhexagon_2| 
                       + |\varhexagon_2 \rangle \langle \varhexagon_1| ],
\end{eqnarray}
where $|\varhexagon_1 \rangle $ and $|\varhexagon_2 \rangle$ refer to the 
two alternating dimer coverings on the elementary hexagons of the dual 
honeycomb lattice (see Fig.~\ref{fig3}). This quantum dimer model is known 
to have a plaquette dimer order by breaking the lattice translation symmetry, 
and the unit cell has been tripled~\cite{PhysRevB.64.144416}. Returning back to the spin language, 
this plaquette dimer order corresponds 
to the quantum plaquette order where in the resonating hexagon with a blue 
cirlce (see Fig.~\ref{fig4}) the ground state can be approximated as
$ ({|{\uparrow \downarrow \uparrow \downarrow \uparrow \downarrow } \rangle} 
+  {|{\downarrow \uparrow \downarrow \uparrow \downarrow \uparrow } \rangle})
/\sqrt{2}$. We have listed the spin state for the six spins on the 
resonating hexagon. This is an even cat state of six spins on the hexagon. 

How does one probe this quantum plaquette order? Thermodynamically, there 
should a finite temperature phase transition as one lowers the temperature. 
Moreover, this order breaks the lattice translation, and one should be able
to observe the magnetic Bragg peaks at the wavevectors $(\pm 4\pi/3, 0)$    
where we have set the original lattice constant to unity. This corresponds 
to the momentum points K and K$'$ in Fig.~\ref{fig4}. In this system, only 
$S^z$ is time reversally odd and can be detected from the conventional neutron 
scattering measurements. So, we expect the magnetic Bragg peak to be observed 
in the $S^z$-$S^z$ correlator for an inelastic neutron scattering measurement. 
Nuclear magnetic resonance (NMR) measurement can also be a convenient
probe of the number of internal fields that are generated by the enlarged
$S^z$ magnetic unit cell from the quantum plaquette order.

How about the elementary excitations? Again, As only $S^z$-$S^z$ correlator 
is detectable, we analyze the physics content that is contained in this correlator. 
If the magnetic state only has $\langle S^z \rangle \neq 0$, the $S^z$-$S^z$ 
correlator can only detect two-magnon excitations and would not be able to 
observe coherent magnon excitations. Because of the transverse field, 
$\langle S^x \rangle \neq 0$ both on the ordered side and on the disordered side. 
Inside the quantum plaquette ordered state, the $S^z$-$S^z$ correlator will be able to observe the
coherent magnon modes, as the $S^z$ operator creates spin-flipping events 
for the $S^x$ configurations and this corresponds to the coherent magnon 
creation. This is also the underlying reason that we can observe the
coherent magnon excitations for the triangular lattice magnet TmMgGaO$_4$ 
and make a reasonable comparison with the spin wave theory. This could 
persist to the disordered side, although the number of modes will be
restored to the one without the translation symmetry breaking. 

Finally, in contrast to the ordered state from the quantum order by disorder 
for the triangular lattice case~\cite{PhysRevLett.84.4457}, the quantum plaquette order for the Kagome 
system is more complicated as it is a quantum entangled state within the 
enlarged unit cell, and the conventional spin wave theory fails. 
Although this quantum plaquette order would manifest itself as Bragg peaks 
in the neutron scattering measurements, its quantum nature distinguishes itself
from other conventional magnetic orders. The magnetic excitation cannot be captured
well by the spin wave theory that is based on the single site magnetic orders. 
To compute 
the magnetic excitation spectrum inside the quantum plaquette ordered phase 
in the future, one needs to first resolve the local energy states within the 
enlarged unit cell because of the entangled nature of the ground state and 
represent these states/operators with a restructured flavor wave theory.

\section{Non-perturbative treatment} 
\label{sec4}

Having established that our model has a quantum plaquette order 
in the weak intrinsic transverse field limit, we continue to understand
the structure of this phase when the field is large. Certainly, in the 
strong field limit, the ground state is a trivial disordered state 
and all the spins try to align themselves with the transverse field. The
natural questions are whether there exists other competing phase between
these two phases in the two opposite limits. If there is no intermediate
phase, what would be the quantum phase transition between these two phases?
To address these questions, we first provide another perspective of the 
quantum plaquette order and then regard this state as a parent state. 
We will explore the instability of this state and see what states we can 
obtain.

It is well-known that the quantum dimer model on the bipartite lattice is described by the 
compact U(1) lattice gauge theory, so is the dimer model for our dual honeycomb lattice.
For two spatial dimensions, this U(1) lattice gauge theory is confining due to the proliferation 
of the spacetime monopole events. Our quantum plaquette order is essentially the consequence
of the confinement. The role of the transverse field in the quantum plaquette ordered
state is to create the spinon-antispinon pair, allow the spinons to hop, and lower 
the energy of the spinons gradually, although the spinons are actually confined. 
Here the spinon/antispinon refers to the defect triangular plaquette that violates 
the up-up-down condition. Under this picture, eventually the spinon band gap vanishes 
and the spinons are condensed, the resulting phase would be a Higgs phase of the 
compact U(1) lattice gauge theory.

\subsection{Transition between the confinement and the Higgs phases}

To describe the transition out of the quantum plaquette ordered phase, we focus 
on the spinon matter and construct a mean-field description to trace the spinons.
To access the transition, we ignore the confined nature of the spinons in the  
quantum plaquette ordered state. This should be appropriate at the transition 
where the translation symmetry is restored, but is not a good approximation 
in the quantum plaquette ordered phase as the translation symmetry is broken.  
We first recast the microscopic Hamiltonian into the following form, 
\begin{eqnarray}
H = \sum_{\triangle_{\boldsymbol r}} 
\frac{J}{2} \big[ \sum_{i \in \triangle_{\boldsymbol r}} S^z_i  - \frac{1}{2} \big]^2 
- h \sum_{i} S^x_i,
\end{eqnarray}
where the first term takes care of the up-up-down condition from the introduction
of the external field on the dipolar component $S^z$, and ``$\triangle_{\boldsymbol r}$''
refers to the triangular plaquette that is centered at ${\boldsymbol r}$. 
As it is known from Fig.~\ref{fig3},
the centers of the triangular plaquettes on the Kagome lattice form a honeycomb 
lattice. We here
introduce the spinon operators in the spirit similar to the one used in the 
context of pyrochlore spin ice system~\cite{PhysRevB.86.104412}, 
\begin{eqnarray}
&&S^+_i = \Phi^\dagger_{\boldsymbol r}  \Phi^{}_{{\boldsymbol r}'} s^+_{{\boldsymbol r}{\boldsymbol r}' }, \\
&& Q_{\boldsymbol r} = \eta_{\boldsymbol r}[\sum_{i \in \triangle_{\boldsymbol r}} S^z_i  - \frac{1}{2}],
\end{eqnarray}
where $S^x_i = (S^+_i + S^-_i)/2$, and the site $i$ is the shared site of two neighbor
triangular plaquettes at ${\boldsymbol r}$ and ${\boldsymbol r}'$. 
Here we choose ${\boldsymbol r}$ to be in the I sublattice of the honeycomb 
lattice and ${\boldsymbol r}'$ 
to be in the II sublattice of the honeycomb lattice, 
$s^+_{{\boldsymbol r}{\boldsymbol r}' }$ is the U(1) 
gauge link variable, and ${Q}_{\boldsymbol r}$ counts the spinon number density 
with $\eta_{\boldsymbol r} = \pm 1$ for I/II sublattice. $\Phi^\dagger_{\boldsymbol r}$
($\Phi_{\boldsymbol r}$) is the creation (annihilation) operator for the spinon
at ${\boldsymbol r}$.
 We have the commutation 
relations
\begin{eqnarray}
&& [\Phi_{\boldsymbol r}, Q_{{\boldsymbol r}'}] = \Phi_{\boldsymbol r} 
                                    \delta_{{\boldsymbol r}{\boldsymbol r}'} \\
&& [\Phi^\dagger_{\boldsymbol r}, Q_{{\boldsymbol r}'}] 
   = - \Phi^\dagger_{\boldsymbol r} \delta_{{\boldsymbol r}{\boldsymbol r}'}.
\end{eqnarray}
Under the parton construction, the physical Hilbert space of the spins is enlarged
to the ones by $\Phi_{\boldsymbol r}$, $Q_{\boldsymbol r}$ and the gauge link.
Once the Hilbert space constraint is imposed, the physical Hilbert space is
restored~\cite{PhysRevB.86.104412}. With this transformation, the physical Hamiltonian can be expressed as 
\begin{eqnarray}
H =   \sum_{\boldsymbol r} \frac{J}{2} Q_{\boldsymbol r}^2 
    - \frac{h}{2}\sum_{\langle {\boldsymbol r}{\boldsymbol r}' \rangle}
       [ \Phi^\dagger_{\boldsymbol r} \Phi^{}_{{\boldsymbol r}'} 
         s^{+}_{{\boldsymbol r}{\boldsymbol r}'} + h.c.]. 
\end{eqnarray}
This model then describes the spinon hopping on the dual honeycomb lattice 
that is minimally coupled with the U(1) gauge link. The first term of the 
above Hamiltonian is the energy penalty that constrains the spinon density
fluctuations. To solve this model, we carry out the standard gauge mean-field
treatment and set ${\Phi_{\boldsymbol r} = e^{-\phi_{\boldsymbol r}}}$ with
$[\phi_{\boldsymbol r}, Q_{{\boldsymbol r}'}] = i \delta_{{\boldsymbol r}{\boldsymbol r}'}$. 
From the knowledge of the previous sixth order perturbation calculation, the system would 
prefer a zero-flux sector for the spinons, and we can choose 
${s^{\pm}_{{\boldsymbol r}{\boldsymbol r}'} =1/2}$ to fix the U(1) gauge link. 
Then, under the coherent state path integral for the bosonic spinons, the 
dispersions of the spinons can be established and are given by
\begin{eqnarray}
\omega_{\pm} ({\boldsymbol k}) = \Big[2J \big( \lambda \pm \frac{h}{4} 
|\sum_{\{ {\boldsymbol b}_i \} } e^{i{\boldsymbol k}\cdot {\boldsymbol b}_i}|  
\big) \Big]^{\frac{1}{2}},
\end{eqnarray}
where $\lambda$ is the lagrangian multiplier to fix the unimodular condition
for ${|\Phi_{\boldsymbol r}| =1}$ and the spinon is condensed at the $\Gamma$ point. 
In addition, $\{ {\boldsymbol b}_i \}$ is the set of three nearest-neighbor bonds
of the dual honeycomb lattice. As the spinon is condensed at the $\Gamma$ point, 
the resulting spin state is a disordered 
state with a finite and uniform $\langle S^x \rangle$, {\sl i.e.}
\begin{eqnarray}
\langle S^x_i \rangle = \frac{1}{2} \big[ s^+_{{\boldsymbol r}{\boldsymbol r}'} 
\langle \Phi^\dagger_{\boldsymbol r} \rangle 
\langle \Phi^{}_{{\boldsymbol r}'}   \rangle
+ h.c.\big],
\end{eqnarray}
where the expectation is taken with respect to the spinon condensate at the $\Gamma$ point. 
The above description would suggest a direct transition from the confining phase to the 
Higgs phase, and the numerical calculation finds the critical point occurs at ${h_c = 1.26J}$. 
This transition is continuous at the mean-field level and may become weakly first order 
when the low-energy gauge fluctuation is included~\cite{PhysRevLett.32.292}.

\begin{figure}[t]
\centering
\includegraphics[width=8.4cm]{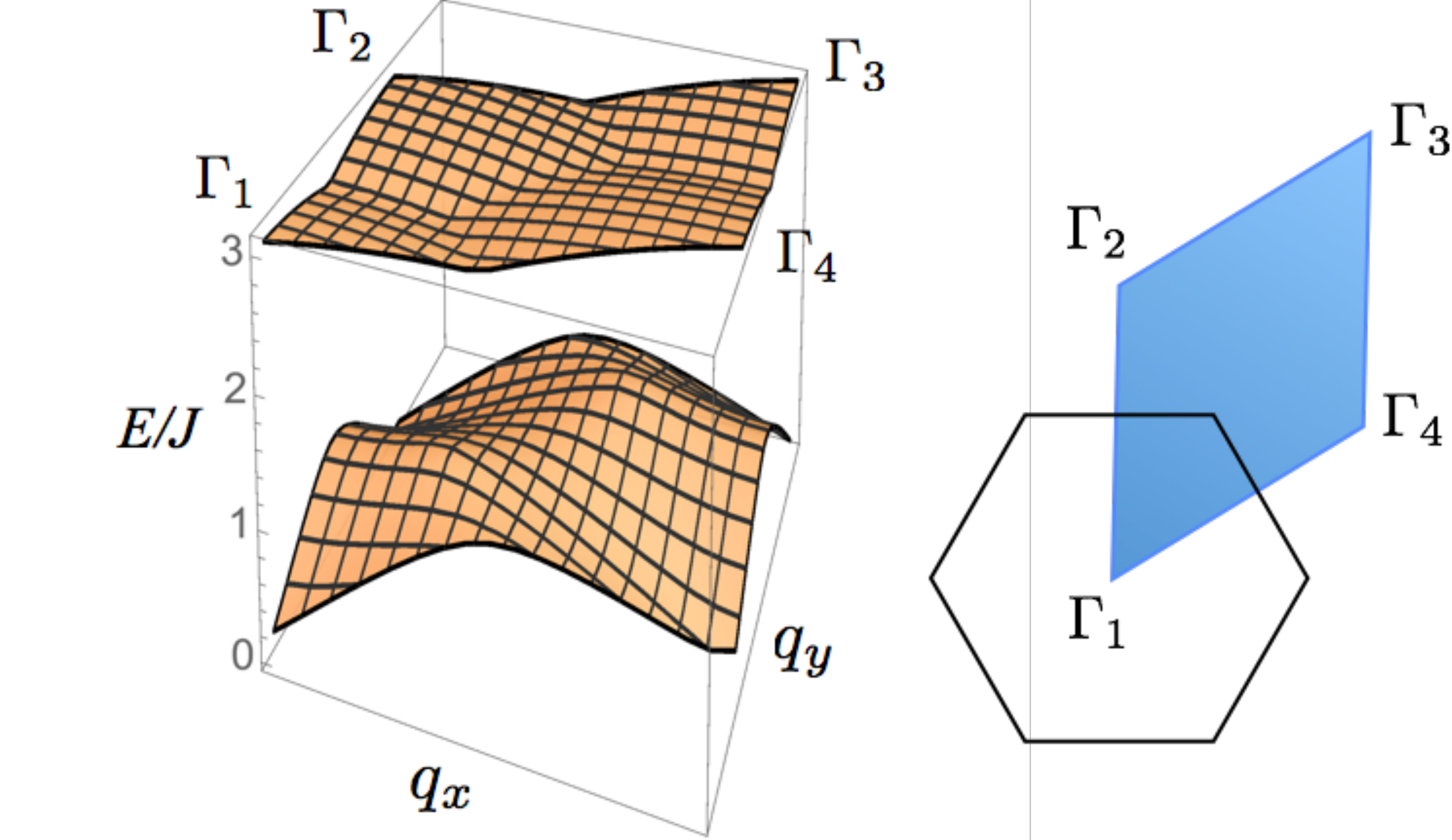}
\caption{The lower and upper excitation edges of the 
2-spinon continuum at the phase transition. We plot the excitation continuum 
with respect to the parallelogram-shaped Brillouin zone. The continuum is gapless at
the $\Gamma$ point. The right panel is the Brillouin zone with all four equivalent 
$\Gamma$ points marked. 
} 
\label{fig5}
\end{figure}

\subsection{Fractionalization and continuum at the criticality}

To probe the possible existence of the fractionalization at the transition
or at least above the energy scale where our mean-field theory would work, we 
suggest the meansurement of the $S^+$-$S^-$ spin correlation function. This spin 
correlator contains the information about the spinon dynamics. It corresponds to 
the creation of the spinon-antispinon, the evolution of them in space and time,
and eventually, the annihilation of them. Thus, the $S^+$-$S^-$ correlator would 
detect the 2-spinon continuum. Although this correlator is not directly measurable 
in the inelastic neutron scattering measurement for the non-Kramers doublet
local moments, as this model does not have the sign problem for the 
quantum Monte Carlo simulation, it can be useful to probe this correlator in the 
future numerical simulation. For the triangular lattice, this model 
has been extensively simulated~\cite{PhysRevLett.84.4457,
PhysRevB.63.224401,PhysRevB.93.235103,PhysRevB.97.085114} 
where the dynamical correlation properties and effect of long-range dipolar interaction were 
numerically studied recently~\cite{2019arXiv190708173L,Koziol2019}. On
the triangular lattice, the system does not have exotic ground state
and the physics is well captured within the conventional symmetry breaking~\cite{PhysRevB.63.224401,PhysRevB.93.235103} and the associated spin wave like quasiparticle picture~\cite{YaoshenTMGO}. 
In contrast, for our Kagome system, we would encounter fractionalization~\cite{PhysRevLett.108.137201,PhysRevB.86.134408,doi:10.1146/annurev-conmatphys-030212-184215}. 
Theoretically, one could obtain the structure of the 
2-spinon continuum within the mean-field approach in the previous section. 
This is established by the energy-momentum conservation with
\begin{eqnarray}
&& {\boldsymbol q} = {\boldsymbol k}_1  + {\boldsymbol k}_2  , \\
&& E_{\boldsymbol q} = \omega_{\mu} ({\boldsymbol k}_1) + \omega_{\nu} ({\boldsymbol k}_2) ,
\end{eqnarray}
where ${\boldsymbol q}$ is the total crystal momentum and $E_{\boldsymbol q}$ 
is the total energy of the two spinons. Here $\mu$ and $\nu$ are the branch 
indices for the spinons and take $\pm$. The minima and maxima of $E_{\boldsymbol q}$ 
define the lower and the upper excitation edges of the 2-spinon continuum. 
The lower and upper edges are plotted in Fig.~\ref{fig5} using the mean-field 
parameters at the transition, and the continuum covers a large energy bandwidth. 
The mean-field theory qualitatively captures the fractionalized nature of the 
excitations at the transition.

\section{Physical properties of the disordered phase}
\label{sec5}

When the transverse field is large, the system will be in a disordered state 
with the Ising spin language. For the model in the absence of the external 
field, the transverse field Ising model on the Kagome lattice remains disordered
for all parameter range. From the experimental point of view, it would be 
quite useful to extract the parameters in the model. For this purpose, as 
we show below, the conventional thermodynamic measurements would be sufficient. 
Moreover, the unique multipolar structure of the local moments generates
peculiar structures in the dynamic spin structure factor measurements.

\subsection{Thermodynamic properties}

As the external magnetic field only couples to the dipole component $S^z$, we now
view the $B$-field as a probing field. The magnetization is non-vanishing only along 
the $z$ direction. Due to the intrinsic transverse field, the 
model does not have any continuous spin rotational symmetry. The magnetic 
susceptibility should simply be a constant in the zero temperature limit. This
constant magnetic susceptibility can be obtained by a conventional self-consistent 
mean-field treatment with
\begin{eqnarray}
H & = & \sum_{\langle ij \rangle} J S^z_i S^z_j - h\sum_i S^x_i - B \sum_i S^z_i \nonumber \\
&\rightarrow & \sum_{\langle ij \rangle} J S^z_i \langle S^z_j\rangle - h\sum_i S^x_i - B \sum_i S^z_i 
\end{eqnarray}
where the expectation is taken with respect to the mean-field ground state. 
From the induced magnetization, it is ready to obtain the zero-temperature magnetic
susceptibility, 
\begin{eqnarray}
\chi_0  =\frac{1}{N} \sum_i\frac{\partial \langle S^z_i \rangle}{\partial B}  
|_{B\rightarrow 0} =\frac{1}{2h+4J},
\end{eqnarray}
where $N$ is the total number of spins. 
Besides the susceptibility in the zero-temperature limit, the Curie-Weiss temperature 
from the high-temperature magnetic susceptibility provides another quantitative 
information with ${\Theta^z_{\text{CW}} = - J}$ where only the $z$ component is meaningful.
With $\chi_0$ and $\Theta^z_{\text{CW}}$, it is sufficient to extract the couplings.

\subsection{Dynamic properties}

The magnetic excitations are measured by inelastic neutron scattering
through the $S^z$-$S^z$ correlation. Here only $S^z$-$S^z$ correlation 
is contained in the inelastic neutron scattering spectrum because only
$S^z$ is coupled to the neutron spin. In the disordered state, $\langle S^x \rangle$
is non-vanishing. As $S^z$ does not commute with $S^x$, what $S^z$ does 
is to flip $S^x$ and create coherent excitations. Thus $S^z$-$S^z$
correlation measures the coherent excitations with respect to the 
disordered state~. This result really arises from the multipolar 
nature of the local moment. As the disordered state is smoothly
connected to the finite temperature paramagnetic state, the coherent
excitations, that are recorded in the $S^z$-$S^z$ correlation,
would persist to the finite temperatures. Experimentally, this may be
mysterious. 

\begin{figure}[t]
\centering
\includegraphics[width=7cm]{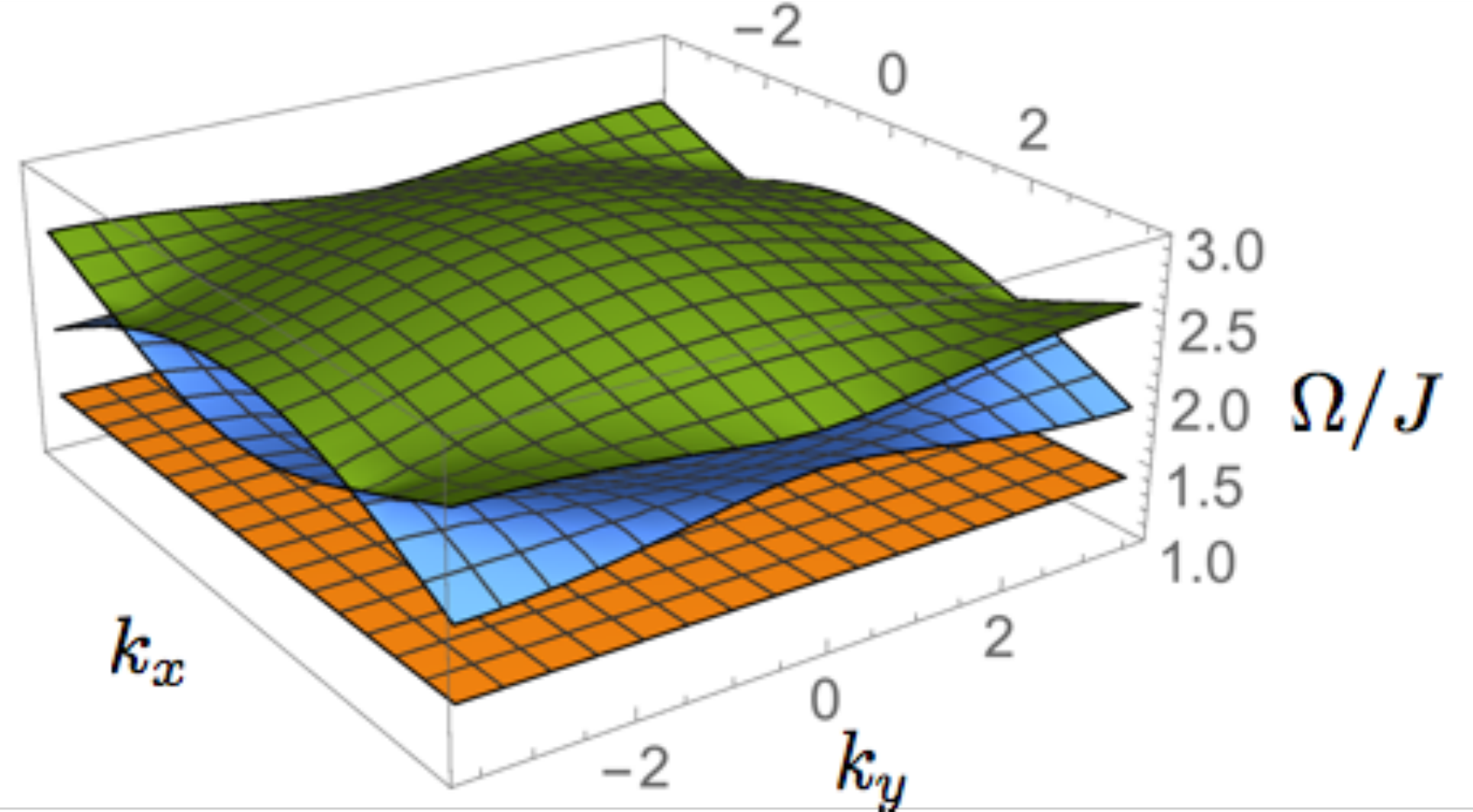}
\caption{The dispersions of the three magnetic excitations in the disordered phase. 
Here we set $h/J=2$. The lattice constant is set to unity. 
} 
\label{fig6}
\end{figure}

We consider the model with ${B=0}$, and set the spin wave transformation as
\begin{eqnarray}
S^z_i &=& \frac{1}{2i} (b_i^{} - b_i^\dagger), \\
S^x_i &=& \frac{1}{2} - b^\dagger_{i} b^{}_i .
\end{eqnarray}
The linear spin wave Hamiltonian is then given as 
\begin{eqnarray}
H_{sw} &=&\sum_{{\boldsymbol k}} \sum_{\mu} h\, 
b^\dagger_{{\boldsymbol k}\mu} b^{}_{{\boldsymbol k}\mu} + 
\sum_{{\boldsymbol k}}\sum_{\mu \neq \nu} A_{\mu\nu}^{} ({\boldsymbol k}) 
b^\dagger_{{\boldsymbol k}\mu}b^{}_{{\boldsymbol k}\nu} 
\nonumber \\
&& + \sum_{{\boldsymbol k}}\sum_{\mu\neq \nu} [ B_{\mu\nu}^{} ({\boldsymbol k})
b^\dagger_{{\boldsymbol k}\mu}b^{\dagger}_{-{\boldsymbol k}\nu}  + h.c. ],
\end{eqnarray}
where $\mu$ refers to the sublattices of the Kagome lattice, and we have 
\begin{eqnarray}
A_{\mu\nu} ({\boldsymbol k}) &=& \frac{J}{2} \cos ({\boldsymbol k}\cdot{\boldsymbol a}_{\mu\nu}) ,
\\
B_{\mu\nu} ({\boldsymbol k}) &=& - \frac{J}{4} \cos ({\boldsymbol k}\cdot{\boldsymbol a}_{\mu\nu}) .
\end{eqnarray}
Here ${\boldsymbol a}_{\mu\nu}$ is the nearest neighbor vector connecting 
sublattice $\mu$ and sublattice $\nu$ for the convention $\mu > \nu$.. 
The magnetic excitations have three branches with the dispersions,
\begin{eqnarray}
\Omega_1 ({\boldsymbol k}) &=& (h^2 -h J)^{\frac{1}{2}} , \nonumber
\\
\Omega_2 ({\boldsymbol k}) &=&  
\Big[
\frac{2h^2 + hJ- hJ [{3+ 2 \sum_{{\boldsymbol a}_{\mu\nu}}  
\cos ( 2{\boldsymbol k}\cdot{\boldsymbol a}_{\mu\nu})}]^{\frac{1}{2}}}{2} 
\Big]^{\frac{1}{2}} , \nonumber
\\
\Omega_3 ({\boldsymbol k}) &=& \Big[
\frac{2h^2 + hJ + hJ [{3+ 2 \sum_{{\boldsymbol a}_{\mu\nu}}  
\cos ( 2{\boldsymbol k}\cdot{\boldsymbol a}_{\mu\nu})}]^{\frac{1}{2}}}{2} 
\Big]^{\frac{1}{2}} , \nonumber 
\end{eqnarray}
where the first band is a flat band. The dispersions are plotted in Fig.~\ref{fig6}. 
The flat band here is not due to the frustration of the interaction and the lattice,
but is from the fact that the exchange part only involves the exchange/hopping
of $b_i^{} - b_i^\dagger$ and does not involve $b_i^{} + b_i^\dagger$. The latter
is the origin of the flat dispersion.

\section{Discussion}
\label{sec6}

In this discussion section, we will do three things. We first review the tripod 
Kagome magnets with the rare-earth ions, and then discuss the Kagome magnets 
from the rare-earth pyrochlore magnets via the dimensional reduction, and finally 
provide some perspectives about the transverse field Ising magnets on the frustrated lattices.

Most of the research about the tripod Kagome magnets are
experimental~\cite{PhysRevB.95.104439,PhysRevLett.116.157201,PhysRevB.95.104439,2018arXiv180604081D}. This 
is a bit different from the Kagome lattice Heisenberg-like magnets 
such as the Herbertsmithite families where both theories and experiments
are quite active~\cite{RevModPhys.88.041002}. 
It is thus beneficial to this new topic of Kagome magnets
if the theoretical inputs are provided. Due to some similiarity 
with the rare-earth pyrochlore and the rare-earth triangular lattice magnets, 
it would be interesting to explore the generic anisotropic spin model on the  
Kagome lattice. The rare-earth tripod Kagome magnets can be another platform  
to study the interplay between the spin-orbit entanglement and the geometrical
frustration. Because this is a $2d$ lattice with relatively lower symmetries, 
more spin interactions (beyond the ones for pyrochlore and triangular lattices) 
are allowed, and new magnetic orders and phases may thus be stabilized. 
While for the Kramers ion, the above expectation may be a simple and 
natural extension from other lattices. The tripod Kagome magnets,
however, do bring additional features for the non-Kramers ions. 
The low symmetry of the Kagome lattice removes the non-Kramers doublets 
completely and splits them into multiple singlets. The physics 
that we have introduced in this paper is about the magnetically 
active lowest two singlets that can be approximately treated 
as an effective pseudospin-1/2 non-Kramers doublets with an
intrinsic transverse field. The quantum plaquette
orders and the phase transitions could be potentially tested 
in future experiments. Moreover, as along as the Ising spin
condition is maintained, even in the presence of the weak transverse 
spin exchange interactions that could exist in real materials, 
our results in this paper will still hold.

It is well-known that, the Kagome lattice magnets can also be
obtained from the rare-earth pyrochlore magnets by applying an 
external magnetic field along the [111] direction~\cite{2007NatPhys566F}. 
Due to the anisotropic coupling to the external magnetic field, 
one sublattice will be polarized. If the pyrochlore system is 
in the spin ice regime, the reduced Kagome system would be in the 
Kagome ice regime. The original non-Kramers doublets of the pyrochlore 
magnets will remain to be non-Kramers doublets under this setting,
and thus there is no intrinsic transverse field here. This, however, may
not be the end of the story. To generate the intrinsic transverse field,
one could grow the pyrochlore thin film along the [111] direction and place it 
on a substrate. The strain from the substrate will modify the lattice symmetry 
and remove the two-fold degeneracy of the non-Kramers doublet. 
On the other hand, the system has the magneto-elastic coupling. This 
coupling was suggested by D. Khomskii to induce the electric dipole moment 
once the spin configuration is modified from the spin ice one~\cite{Khomskii}. 
The distortion of the lattice and/or the displacement of the ions will 
necessarily lower the lattice symmetry and generate a finite 
splitting among the non-Kramers doublet, and this can be treated 
as an intrinsic transverse field.

In this paper, we have delivered the frustrated quantum Ising model  
with an intrinsic transverse field on the Kagome lattice. Since 
the mechanism for the intrinsic transverse field with low crystal 
symmetries can generally apply to the non-Kramers ions, the quantum 
Ising model can be realized and explored among other frustrated 
rare-earth magnets such as the FCC double perovskites~\cite{PhysRevB.95.085132}.

\emph{Acknowledgments.}---We acknowledge Zhiling Dun, 
Martin Mourigal, Joseph Paddison and Tao Xiang for discussion,
and Cenke Xu and Yang Qi for a more recent discussion. 
This work is supported by the Ministry of Science and Technology 
of China with Grant No.2016YFA0301001, 2016YFA0300501, 2018YFE0103200 
and by the General Research Fund (GRF) No.17303819 from the Research 
Grants Council of Hong Kong. This work was performed in part at Aspen 
Center for Physics, which is supported by National Science Foundation 
grant PHY-1607611. This work was partially supported by a grant from 
the Simons Foundation.


\bibliography{refs}

\end{document}